\documentclass[12pt]{article}
\usepackage{a4wide}
\usepackage{latexsym}
\usepackage{amsmath}
\usepackage{amsfonts}
\usepackage{amscd}
\usepackage{cite}
\usepackage{placeins}

\usepackage{pslatex}
\usepackage{graphicx}
\usepackage[latin1,utf8]{inputenc}
\usepackage[T1]{fontenc}

\allowdisplaybreaks

\newcommand{\bq}{\begin{eqnarray}}
\newcommand{\eq}{\end{eqnarray}}
\newcommand{\eps}{\varepsilon}

\newcommand{\xp}{x'}

\begin{document}

\thispagestyle{empty}

\begin{flushright}
  MITP/21-014
% \\ version of \today
\end{flushright}

\vspace{1.5cm}

\begin{center}
  {\Large\bf The H-graph with equal masses in terms of multiple polylogarithms\\
  }
  \vspace{1cm}
  {\large Philipp Alexander Kreer and Stefan Weinzierl \\
  \vspace{1cm}
      {\small \em PRISMA Cluster of Excellence, Institut f{\"u}r Physik, }\\
      {\small \em Johannes Gutenberg-Universit{\"a}t Mainz,}\\
      {\small \em D - 55099 Mainz, Germany}\\
  } 
\end{center}

\vspace{2cm}

% abstract ---------------------------------------
\begin{abstract}\noindent
  {
The initial phase of the inspiral process of a binary system producing gravitational waves
can be described by perturbation theory.
At the third post-Minkowskian order a two-loop double box graph, known
as H-graph contributes. 
We consider the case where the two objects making up the binary system have equal masses.
We express all master integrals  
related to the equal-mass H-graph up to weight four in terms of multiple polylogarithms.
We provide a numerical program which evaluates all master integrals up to weight four 
in the physical regions with arbitrary precision.
   }
\end{abstract}

\vspace*{\fill}

% -----------------------------------------------------------------------------
\newpage

\section{Introduction}
\label{sect:intro}

The initial phase of the inspiral process of a binary system producing gravitational waves
can be described by 
perturbation theory \cite{Buonanno:1998gg,Buonanno:2000ef,Damour:2012mv,Damour:2017ced,Bini:2020nsb,Bini:2020hmy}.
Effective field theory methods \cite{Goldberger:2004jt,Cheung:2018wkq,Porto:2016pyg,Levi:2018nxp} provide a link
between general relativity and particle physics.
There has been a fruitful interplay between these communities 
in recent years \cite{Foffa:2016rgu,Foffa:2019hrb,Bini:2020uiq,Bini:2020rzn,Bjerrum-Bohr:2018xdl,Cristofoli:2019neg,Kosower:2018adc,Bern:2019nnu,Bern:2019crd,Bern:2021dqo,Blumlein:2019zku,Blumlein:2019bqq,Blumlein:2020pog,Blumlein:2020znm,Blumlein:2020pyo,Blumlein:2021txj,Foffa:2019rdf,Foffa:2019yfl,Kalin:2019rwq,Kalin:2019inp,Kalin:2020mvi,Kalin:2020fhe,Liu:2021zxr,Herrmann:2021lqe,DiVecchia:2021bdo,Bjerrum-Bohr:2021vuf}.
At the third post-Minkowskian order a two-loop double box graph, known
as H-graph contributes. 
This is the most complicated graph entering the third post-Minkowskian order.
The H-graph is shown in fig.~\ref{fig_H-graph}, 
where solid lines represent the two massive objects making up the binary system and dashed lines represent gravitons.
In this article we consider the case where the two massive objects have equal masses, the more general case of unequal masses will
be considered in a subsequent publication.
We consider the H-graph in the relativistic setting without any non-relativistic approximations.

The H-graph with equal masses was first studied in ref.~\cite{Bianchi:2016yiq} in the context of quantum chromodynamics
(where the solid lines represent massive quarks and the dashed lines gluons).
In ref.~\cite{Bianchi:2016yiq} a set of canonical master integrals and the differential equation for these master integrals was derived.
The differential equation is in $\eps \; d\log$-form (where $\eps$ denotes the dimensional regularisation parameter), 
however the arguments of the various $d\log$'s contain several square roots.
It is therefore not evident, if all master integrals can be expressed in terms of multiple polylogarithms or not.

In this article we present for all master integrals results up to weight four in terms of multiple polylogarithms.
The challenge is not to express the top-level master integral with propagators all to power one up to weight four in terms of multiple polylogarithms.
This particular integral is up to weight four rather simple and the result in terms of multiple polylogarithms is given in ref.~\cite{Bianchi:2016yiq}.
What is not known and more challenging, are the analytic expressions of all master integrals up to weight four.
This concerns in particular the remaining master integrals in the top-level sector and a few master integrals from sub-sectors.

Expressing all master integrals in terms of multiple polylogarithms would be straightforward 
if all arguments of the $d\log$'s can be rationalised simultaneously.
In the present case we expect that a transformation which simultaneously rationalises all square roots does not exist.
However, the fact that not all roots can be rationalised simultaneously does not necessarily imply that the Feynman integrals
cannot be expressed in terms of multiple polylogarithms, as shown for the first time in ref.~\cite{Heller:2019gkq}.
By combining different techniques we are able to express all master integrals up to weight four in terms of multiple polylogarithms.
It turns out that we may rationalise simultaneously all square roots except one.
The square root, which cannot be rationalised in combination with the other square roots, appears up to weight four only in one master integral.
This master integral can be evaluated in the Feynman parameter representation to multiple polylogarithms.
For all other master integrals we use the method of differential equations together with a rationalisation of the square roots.

We provide the results for the master integrals in electronic form in two ways:
On the one hand, we provide symbolic expressions in terms of multiple polylogarithms for all master integrals
up to weight four.
On the other hand, we provide a numerical program, which evaluates all master integrals up to weight four
at a given kinematic point inside the physical region
with a user-defined precision.
We also would like to mention in passing ``Loopedia'' \cite{Bogner:2017xhp} as a usufull database for Feynman integrals.

This article is organised as follows:
In section~\ref{sect:notation} we introduce our notation.
The master integrals are defined in section~\ref{sect:masters}.
The differential equation for the master integrals is given in section~\ref{sect:differential_equation}.
The method for the solution in terms of multiple polylogarithms is  discussed in section~\ref{sect:solution}.
The results are presented in section~\ref{sect:results}.
Finally, our conclusions are given in section~\ref{sect:conclusions}.
Appendix~\ref{sect:supplement} describes in detail the electronic file attached to the arxiv version of this article,
containing our results in electronic form.

% -----------------------------------------------------------------------------

\section{Notation}
\label{sect:notation}

\begin{figure}
\begin{center}
\includegraphics[scale=1.0]{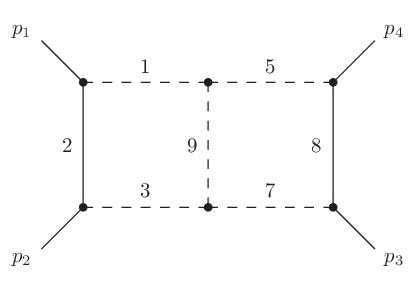}
\end{center}
\caption{
The H-graph. Solid lines denote massive objects of mass $m$, dashed lines denote massless particles.
}
\label{fig_H-graph}
\end{figure}

\begin{figure}
\begin{center}
\includegraphics[scale=1.0]{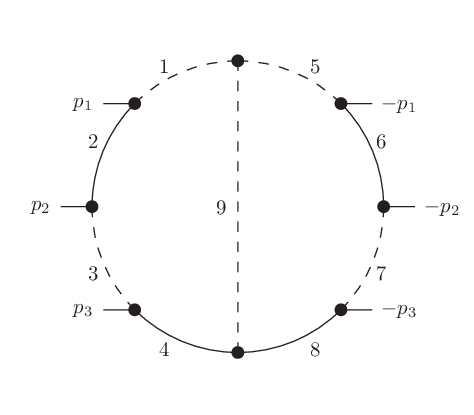}
\end{center}
\caption{
The auxiliary graph with nine propagators.
Solid lines denote massive objects of mass $m$, dashed lines denote massless particles.
}
\label{fig_auxiliary-graph}
\end{figure}

We are interested in the H-graph shown in fig.~\ref{fig_H-graph}.
Solid lines denote massive objects of mass $m$, dashed lines denote massless particles.
In the application towards binary systems the two solid lines correspond to the two objects making up the binary system, 
the massless particles to gravitons. 
The name H-graph stems from the fact that the gravitons form the letter ``H'' 
(which in our figure is rotated by $90^\circ$).
We may express any Feynman integral with non-trivial numerators in terms of scalar integrals and hence it is sufficient to
focus on scalar integrals.
The H-graph has seven propagators (labelled $1,2,3,5,7,8,9$ in fig.~\ref{fig_H-graph}).
In order to express any scalar product involving the loop momenta as a linear combination of inverse
propagators we have to consider an auxiliary graph with nine propagators shown in fig.~\ref{fig_auxiliary-graph}.
Hence, we consider the integrals
\bq
\label{def_integral}
 I_{\nu_1 \nu_2 \nu_3 \nu_4 \nu_5 \nu_6 \nu_7 \nu_8 \nu_9}
 & = &
 e^{2 \gamma_E \eps}
 \left(\mu^2\right)^{\nu-D}
 \int \frac{d^Dk_1}{i \pi^{\frac{D}{2}}} \frac{d^Dk_2}{i \pi^{\frac{D}{2}}} 
 \prod\limits_{j=1}^9 \frac{1}{ \left(P_j\right)^{\nu_j} },
\eq
where $D=4-2\eps$ denotes the number of space-time dimensions,
$\gamma_E$ denotes the Euler-Mascheroni constant, 
$\mu$ is an arbitrary scale introduced to render the Feynman integral dimensionless,
and the quantity $\nu$ is defined by
\bq
 \nu & = &
 \sum\limits_{j=1}^9 \nu_j.
\eq
The inverse propagators $P_j$ are defined as follows:
\begin{align}
 P_1 & = -\left(k_1+p_1\right)^2,
 &
 P_2 & = -k_1^2 + m^2,
 &
 P_3 & = -\left(k_1-p_2\right)^2,
 \nonumber \\
 P_4 & = -\left(k_1-p_2-p_3\right)^2 + m^2,
 &
 P_5 & = -\left(k_2+p_1\right)^2,
 &
 P_6 & = -k_2^2 + m^2,
 \nonumber \\
 P_7 & = -\left(k_2-p_2\right)^2,
 & 
 P_8 & = -\left(k_2-p_2-p_3\right)^2 + m^2,
 &
 P_9 & = -\left(k_1-k_2\right)^2.
\end{align}
The external momenta satisfy
\bq
 p_1^2 \; = \; p_2^2 \; = \; p_3^2 \; = \; p_4^2 \; = \; m^2.
\eq
The Mandelstam variables are defined by
\bq
 s \; = \; \left(p_1+p_2\right)^2,
 \;\;\;\;\;\;
 t \; = \; \left(p_2+p_3\right)^2,
 \;\;\;\;\;\;
 u \; = \; \left(p_1+p_3\right)^2,
\eq
and satisfy
\bq
 s + t + u & = & 4 m^2.
\eq
We are interested in the integrals with $\nu_4 \le 0$ and $\nu_6 \le 0$.
With the help of integration-by-parts identities \cite{Tkachov:1981wb,Chetyrkin:1981qh}
implemented in public available computer programs 
\cite{Smirnov:2019qkx,vonManteuffel:2012np,Klappert:2020nbg,Lee:2013mka}
we may reduce 
all integrals $I_{\nu_1 \nu_2 \nu_3 \nu_4 \nu_5 \nu_6 \nu_7 \nu_8 \nu_9}$
with $\nu_4 \le 0$ and $\nu_6 \le 0$ to linear combinations of $25$ master integrals.
Thus we only need to compute $25$ master integrals.

The essential complication in the computation of the master integrals is the occurrence of square roots.
We encounter the following square roots:
\bq
\label{def_roots}
 r_1
 & = &
 \sqrt{-s^{\mathstrut}}\sqrt{4 m^2-s},
 \nonumber \\
 r_2
 & = &
 \sqrt{-t^{\mathstrut}}\sqrt{4 m^2-t},
 \nonumber \\
 r_3 
 & = &  
 \sqrt{-s\vphantom{\left(m^2-t\right)^2}}\sqrt{4m^6-s\left(m^2-t\right)^2},
 \nonumber \\
 r_4
 & = &
 \sqrt{-s-t^{\mathstrut}}\sqrt{4 m^2 - s - t}.
\eq
Note that we write
\bq
 \sqrt{-s^{\mathstrut}}\sqrt{4 m^2-s}
 & \mbox{and not} &
 \sqrt{-s\left(4m^2-s\right)}.
\eq
In the Euclidean region ($s<0, t<0, m^2>0$) the arguments of all roots are positive and the two forms
are equivalent.
In regions where $s>0$ or $t>0$ we
we have to add a small imaginary part according to
Feynman's $i\delta$ prescription ($s \rightarrow s+i\delta, t \rightarrow t+i\delta$ with $\delta>0$) and
the two forms may differ.
An example is provided by $s=25$ and $m^2=4$.
We have (with the standard choice of the branch cut of the square root along the negative real axis)
\bq
 \left. \sqrt{-s^{\mathstrut}-i\delta}\sqrt{4 m^2-s-i\delta}\right|_{s=25,m^2=4,\delta \rightarrow 0+}
 & = & - 15,
 \nonumber \\
 \left. \sqrt{\left(-s-i\delta\right)\left(4 m^2-s-i\delta\right)}\right|_{s=25,m^2=4,\delta \rightarrow 0+}
 & = & + 15.
\eq
The form in eq.~(\ref{def_roots}) simplifies the analytic continuation from the Euclidean region to the physical
region of interest.

% -----------------------------------------------------------------------------

\section{Master integrals}
\label{sect:masters}

We recall that for the number of space-time dimensions we set
$D=4-2\eps$.
As master integrals we use
\bq
\label{def_master_integrals}
 J_{1}
 & = &
 \eps^2 I_{020000020},
 \nonumber \\
 J_{2}
 & = &
 \eps^2 \frac{\left(-s\right)}{\mu^2} I_{020020100},
 \nonumber \\
 J_{3}
 & = &
 \frac{\eps^2 \left(1+4\eps\right)}{\left(1+\eps\right)} \frac{m^2}{\mu^2} I_{010020002},
 \nonumber \\
 J_{4}
 & = &
 \eps^2 \frac{\left(-s\right)}{\mu^2} I_{001020002},
 \nonumber \\
 J_{5}
 & = &
 \frac{1}{2} \eps^2 \frac{r_2}{\mu^2} \left( 2 I_{020000012} + I_{020000021} \right),
 \nonumber \\
 J_{6}
 & = &
 \eps^2 \frac{\left(-t\right)}{\mu^2} I_{020000021},
 \nonumber \\
 J_{7}
 & = &
 \eps^2 \frac{\left(-s\right)^2}{\mu^4} I_{201020100},
 \nonumber \\
 J_{8}
 & = &
 \eps^3 \frac{r_1}{\mu^2} I_{111000020},
 \nonumber \\
 J_{9}
 & = &
 2 \eps^3 \frac{r_1}{\mu^2} I_{011020001},
 \nonumber \\
 J_{10}
 & = &
 \eps^3 \frac{r_1}{\mu^2} I_{020010101},
 \nonumber \\
 J_{11}
 & = &
 \eps^2 \frac{m^2 r_1}{\mu^4} I_{030010101},
 \nonumber \\
 J_{12}
 & = &
 \eps^2 \frac{\left(-s\right)}{\mu^2}
 \left[ 
        \frac{3}{2} \eps I_{020010101} + \frac{m^2}{\mu^2} I_{020020101} - \frac{m^2}{\mu^2} I_{030010101}
 \right],
 \nonumber \\
 J_{13}
 & = &
 2 \eps^3 \frac{r_2}{\mu^2} I_{110000021},
 \nonumber \\
 J_{14}
 & = &
 \eps^3 \frac{\left(-s\right) r_1}{\mu^4} I_{111020100},
 \nonumber \\
 J_{15}
 & = &
 \eps^3 \frac{r_3}{\mu^4} I_{111000021},
 \nonumber \\
 J_{16}
 & = &
 \eps^3 \frac{r_1}{\mu^2} \left( I_{111\left(-1\right)00021} - \frac{m^2}{\mu^2} I_{111000021} \right),
 \nonumber \\
 J_{17}
 & = &
 \eps^2 \frac{\left(-s\right) r_2}{\mu^4} \left( \frac{m^2}{\mu^2} I_{111000031} - \eps I_{111000021} \right),
 \nonumber \\
 J_{18}
 & = &
 4 \eps^4 \frac{r_4}{\mu^2} I_{011010011},
 \nonumber \\
 J_{19}
 & = &
 2 \eps^3 \frac{\left(-s\right) r_2}{\mu^4} I_{011010012},
 \nonumber \\
 J_{20}
 & = &
 2 \eps^3 \frac{m^2 r_1}{\mu^4} I_{011010021},
 \nonumber \\
 J_{21}
 & = &
 \frac{1}{2} \eps^2 \frac{\left(-s\right)}{\mu^2} 
  \left[
   2 \frac{m^2}{\mu^2} \left( \frac{m^2}{\mu^2} I_{021010021} - 2 \eps I_{011010021} \right) 
   - \eps \frac{\left( 2 m^2 - t \right)}{\mu^2} I_{011010012}
  \right],
 \nonumber \\
 J_{22}
 & = &
 \eps^4 \frac{\left(-s\right) \left(4m^2-s\right)}{\mu^4} I_{111010110},
 \nonumber \\
 J_{23}
 & = &
 \eps^4 \frac{\left(-s\right)^2 r_2}{\mu^6} I_{111010111},
 \nonumber \\
 J_{24}
 & = &
 \eps^4 \frac{\left(-s\right) r_1}{\mu^4} I_{111\left(-1\right)10111},
 \nonumber \\
 J_{25}
 & = &
 \eps^4 \frac{\left(-s\right)}{\mu^2}
 \left[
       I_{111\left(-1\right)1\left(-1\right)111} 
       + \frac{\left(-s\right)}{\mu^2} I_{111\left(-1\right)10111}
       - \frac{\left(-t\right)}{\mu^2} I_{111010110}
 \right]
 \nonumber \\
 & &
 + \eps^3 \frac{\left(-s\right)}{\mu^2} 
   \left[
     -2 \eps I_{011010011}
     + 2 \frac{m^2}{\mu^2} I_{011010021}
     + 2 I_{111\left(-1\right)00021} - 2 \frac{m^2}{\mu^2} I_{111000021}
 \right. \nonumber \\
 & & \left.
     + \frac{\left(-s\right)}{\mu^2} I_{111020100}
     + I_{020010101}
     - \frac{2}{\eps} \frac{m^2}{\mu^2} I_{030010101}
     - 4 I_{011020001}
     - I_{111000020}
   \right]
 \nonumber \\
 & &
 + \frac{\eps^2}{\left(1-2\eps\right)} \frac{\left(-s\right)}{\mu^2}
   \left[
          - \eps \frac{\left(-s\right)}{\mu^2} I_{201020100}
          + \frac{1}{2} I_{001020002}
   \right].
\eq
This choice coincides with the choice of master integrals in \cite{Bianchi:2016yiq} up to relabelling and trivial prefactors.

% -----------------------------------------------------------------------------

\section{The differential equation}
\label{sect:differential_equation}

We consider the derivatives of the master integrals $J_{1}$-$J_{25}$ with respect 
to the kinematic variables $s, t$ and $m^2$.
The derivatives can again be written as a linear combination of the master integrals.
This gives us the differential equation as
\bq
\label{def_diff_eq}
 d J & = & A J,
\eq
with
\bq
 A & = & A_s \; ds + A_t \; dt + A_{m^2} \; dm^2.
\eq
For the choice of master integrals as in eq.~(\ref{def_master_integrals}), the differential equation is in
$\eps$-form \cite{Henn:2013pwa}
and we write
\bq
\label{def_eps_A}
 A & = &
 \eps \sum\limits_{k=1}^{17} C_k \omega_k,
\eq
where the $C_k$'s are $25 \times 25$-matrices, whose entries are rational numbers.
The $\omega_k$'s are differential one-forms.
They are given by
\bq
\label{def_omega}
 \omega_{1}
 & = & 
 d \ln\left(\frac{-s}{\mu^2}\right),
 \nonumber \\
 \omega_{2}
 & = & 
 d \ln\left(\frac{-t}{\mu^2}\right),
 \nonumber \\
 \omega_{3}
 & = & 
 d \ln\left(\frac{m^2}{\mu^2}\right),
 \nonumber \\
 \omega_{4}
 & = & 
 d \ln\left(\frac{4m^2-s}{\mu^2}\right),
 \nonumber \\
 \omega_{5}
 & = & 
 d \ln\left(\frac{4m^2-t}{\mu^2}\right),
 \nonumber \\
 \omega_{6}
 & = & 
 d \ln\left(\frac{4m^2-s-t}{\mu^2}\right),
 \nonumber \\
 \omega_{7}
 & = & 
 d \ln\left(\frac{-s-t}{\mu^2}\right),
 \nonumber \\
 \omega_{8}
 & = & 
 d \ln\left(\frac{2m^2-s-r_1}{2m^2-s+r_1}\right),
 \nonumber \\
 \omega_{9}
 & = & 
 d \ln\left(\frac{2m^2-t-r_2}{2m^2-t+r_2}\right),
 \nonumber \\
 \omega_{10}
 & = & 
 d \ln\left(\frac{\left(-s\right)\left(m^2-t\right)-r_3}{\left(-s\right)\left(m^2-t\right)+r_3}\right),
 \nonumber \\
 \omega_{11}
 & = & 
 d \ln\left(\frac{2m^2-s-t-r_4}{2m^2-s-t+r_4}\right),
 \nonumber \\
 \omega_{12}
 & = & 
 d \ln\left(\frac{s t-r_1r_2}{s t+r_1r_2}\right),
 \nonumber \\
 \omega_{13}
 & = & 
 d \ln\left(\frac{\left(-s\right)\left(4m^4-m^2 s -st\right)-r_1r_3}{\left(-s\right)\left(4m^4-m^2 s -st\right)+r_1r_3}\right),
 \nonumber \\
 \omega_{14}
 & = & 
 d \ln\left(\frac{p_{14} - q_{14} r_1 r_3}{p_{14} + q_{14} r_1 r_3}\right),
 \nonumber \\
 \omega_{15}
 & = & 
 d \ln\left(\frac{\left(-s\right)\left(t^2-3 m^2 t\right)-r_2r_3}{\left(-s\right)\left(t^2-3 m^2 t\right)+r_2r_3}\right),
 \nonumber \\
 \omega_{16}
 & = & 
 d \ln\left(\frac{p_{16} - q_{16} r_1 r_4}{p_{16} + q_{16} r_1 r_4}\right),
 \nonumber \\
 \omega_{17}
 & = & 
 d \ln\left(\frac{\left(-t\right)\left(4m^2 -t\right)+\left(-s\right)\left(2m^2-t\right)-r_2r_4}{\left(-t\right)\left(4m^2 -t\right)+\left(-s\right)\left(2m^2-t\right)+r_2r_4}\right).
\eq
In the definition of $\omega_{14}$ and $\omega_{16}$ the polynomials $p_{14}, q_{14}, p_{16}$ and $q_{16}$
appear. They are given by
\bq
 p_{14}
 & = &
 \left(-s\right) 
   \left(
          2 m^4 \left( 8 m^4 - 4 m^2 t + t^2 \right) 
          + 2 m^2 \left(-s\right) \left( 4 m^4 - 5 m^2 t + 2 t^2 \right)
          + \left(-s\right)^2 \left(m^2-t\right)^2
   \right),
 \nonumber \\
 q_{14}
 & = &
 2 m^2 \left(2m^2-t\right) + \left(-s\right) \left(m^2-t\right),
 \nonumber \\
 p_{16}
 & = &
 - 8 m^4 t + 2 m^2 t^2 + 10 m^2 s t - 8 s m^2 \left(2 m^2-s\right) - s \left(s+t\right)^2,
 \nonumber \\
 q_{16}
 & = &
 4 m^2-s-t.
\eq
In solving the differential equation we may always keep one variable constant.
A typical choice would be $m^2=\mu^2=\mathrm{const}$.
In this case $\omega_3=0$ and the number of non-zero differential one-forms reduces to $16$, in agreement
with the number reported in ref.~\cite{Bianchi:2016yiq}.
The differential one-forms reported in ref.~\cite{Bianchi:2016yiq} can be written as linear combinations
of the ones defined in eq.~(\ref{def_omega}).
We provide the matrix $A$ in electronic form, see appendix~\ref{sect:supplement}.

The differential equation eq.~(\ref{def_diff_eq})
is easily solved in terms of iterated integrals.
In general, an iterated integrals is defined as follows \cite{Chen}:
Let $M$ be a $n$-dimensional manifold and
\bq
 \gamma & : & \left[a,b\right] \rightarrow M
\eq
a path with start point ${x}_i=\gamma(a)$ and end point ${x}_f=\gamma(b)$.
Suppose further that $\omega_1$, ..., $\omega_r$ are differential $1$-forms on $M$.
Let us write
\bq
 f_j\left(\lambda\right) d\lambda & = & \gamma^\ast \omega_j
\eq
for the pull-backs to the interval $[a,b]$.
For $\lambda \in [a,b]$ the $k$-fold iterated integral 
of $\omega_1$, ..., $\omega_r$ along the path $\gamma$ is defined
by
\bq
 I_{\gamma}\left(\omega_1,...,\omega_r;\lambda\right)
 & = &
 \int\limits_a^{\lambda} d\lambda_1 f_1\left(\lambda_1\right)
 \int\limits_a^{\lambda_1} d\lambda_2 f_2\left(\lambda_2\right)
 ...
 \int\limits_a^{\lambda_{r-1}} d\lambda_r f_r\left(\lambda_r\right).
\eq
Multiple polylogarithms are a special case of iterated integrals, where all pull-back's are of the form
\bq
 \gamma^\ast \omega_j \; = \; f_j\left(\lambda\right) d\lambda \; = \; \frac{d\lambda}{\lambda-z_j}
\eq
for some $z_j \in {\mathbb C}$.
Allowing trailing zeros, we define multiple polylogarithms as $G(z_1,\dots,z_r;\lambda)$ follows:
If all $z$'s are equal to zero, we define $G(z_1,\dots,z_r;\lambda)$ by
\bq
 G(\underbrace{0,\dots,0}_{r-\mathrm{times}};\lambda)
 & = & 
 \frac{1}{r!} \ln^r\left(\lambda\right).
\eq
This definition includes as a trivial case
$G(;\lambda) = 1$.
If at least one variable $z$ is not equal to zero we define recursively
\bq
 G\left(z_1,z_2\dots,z_r;\lambda\right)
 & = &
 \int\limits_0^\lambda
 \frac{d\lambda_1}{\lambda_1-z_1}
 G\left(z_2\dots,z_r;\lambda_1\right).
\eq
The weight of the multiple polylogarithm $G(z_1,\dots,z_r;\lambda)$ is $r$.

We would like to express the master integrals $J_{1}$-$J_{25}$ in terms of multiple polylogarithms.
This would be straightforward if all arguments of the logarithms appearing in eq.~(\ref{def_omega})
would be rational functions in the kinematic variables $s$, $t$ and $m^2$.
The obstruction is given by the occurrence of the square roots $r_1$-$r_4$.
The occurrence of square roots is not always a problem. 
If there is a transformation of the kinematic variables, which simultaneously rationalises all square roots,
we may again easily convert all iterated integrals to multiple polylogarithms.
The challenge we face in converting all iterated integrals to multiple polylogarithms 
is related to the fact that we do not expect such a transformation to exist.
The non-existence of a transformation has been proven in the slightly different case 
of two-loop corrections to the Drell-Yan process \cite{Besier:2019hqd}.
However, the fact that not all roots can be rationalised simultaneously does not necessarily imply that the Feynman integrals
cannot be expressed in terms of multiple polylogarithms, as shown for the first time in ref.~\cite{Heller:2019gkq}.
It only means that the method of differential equations does not lead in a straightforward way to multiple polylogarithms.
Other methods, like direct integration in Feynman parameter space, may produce a result in terms of multiple polylogarithms.

% -----------------------------------------------------------------------------

\section{Solution in terms of multiple polylogarithms}
\label{sect:solution}

In this section we express all master integrals up to weight four in terms of multiple polylogarithms.
Without loss of generality we set
\bq
 \mu^2 & = & m^2.
\eq
We write
\bq
 J_i 
 & = & 
 \sum\limits_{j=0}^\infty J_i^{(j)} \eps^j
\eq
for the expansion in the dimensional regularisation parameter $\eps$ and we compute for each master integral
the coefficients $J_i^{(0)}$-$J_i^{(4)}$.
Up to weight four the root $r_4$ enters only the master integral $J_{18}$, all other master integrals
are independent of the root $r_4$ up to weight four.
The root $r_4$ will enter other master integrals at higher weights.
As the terms up to weight four are the relevant ones for two-loop calculations, we split 
the calculation of the master integrals into two cases:
The first case consists of all master integrals except $J_{18}$, the second case consists of the master integral $J_{18}$.

\subsection{The master integrals except $J_{18}$}

If we restrict our attention to the master integrals $J_{1}-J_{17}$ and $J_{19}-J_{25}$ up to weight four, we only have to
deal with the roots $r_1$, $r_2$ and $r_3$.
These roots can be rationalised simultaneously.
Up to weight four the root $r_3$ appears only in the master integrals $J_{15}-J_{17}$ and $J_{24}-J_{25}$,
the master integrals $J_{1}-J_{14}$ and $J_{19}-J_{23}$ involve up to weight four only the roots $r_1$ and $r_2$.

The roots $r_1$ and $r_2$ are rationalised by the standard transformations
\bq
 s \; = \; - \frac{\left(1-x\right)^2}{x} m^2,
 & &
 t \; = \; - \frac{\left(1-y\right)^2}{y} m^2.
\eq
The value $s=0$ corresponds to $x=1$ (and the value $t=0$ corresponds to $y=1$).
It will be convenient to introduce 
\bq
 \bar{x} \; = \; 1-x,
 & &
 \bar{y} \; = \; 1-y.
\eq
Then $s=0$ corresponds to $\bar{x}=0$ and $t=0$ corresponds to $\bar{y}=0$.
In terms of $\bar{x}$ and $\bar{y}$ we have
\bq
 s \; = \; - \frac{\bar{x}^2}{1-\bar{x}} m^2,
 & &
 \bar{x} \; = \; \frac{s}{2m^2} + \frac{r_1}{2m^2},
 \nonumber \\
 t \; = \; - \frac{\bar{y}^2}{1-\bar{y}} m^2,
 & &
 \bar{y} \; = \; \frac{t}{2m^2} + \frac{r_2}{2m^2}.
\eq
With the help of the methods of refs.~\cite{Besier:2018jen,Besier:2019kco} we find a transformation, which rationalises
in addition $r_3$:
\bq
\label{rationalise_r3}
 \bar{x} 
 & = &
 \frac{4 \xp \left(1-\bar{y}\right) \left(1-\xp + \xp \bar{y} \right)}
      {\left[ 1 + \left(1-\bar{y}+\bar{y}^2\right) \xp \right] \left[ 1 - \left(1-\bar{y}+\bar{y}^2\right) \xp \right]},
 \nonumber \\
 \xp 
 & = &
 \frac{\left(1-\bar{y}\right)\left[2\bar{x} - \left(1-\bar{x}\right) \frac{r_3}{m^4}\right]}{\bar{x}\left[4\left(1-\bar{y}\right)^2-\bar{x} \left(1-\bar{y}+\bar{y}^2\right)^2\right]}.
\eq
The point $\bar{x}=0$ corresponds to $\xp=0$.

We integrate the differential equation from the boundary point $s=0$, $t=0$ (corresponding to $\bar{x}=0$, $\bar{y}=0$).
The boundary values at $s=0$, $t=0$ are obtained from the results of ref.~\cite{Bianchi:2016yiq}.
We first integrate in $\bar{y}$ (on the hypersurface $s=0$).
In a second step we integrate in $\bar{x}$ or $\xp$ (on the hypersurface $t=\mathrm{const}$).

Actually, ref.~\cite{Bianchi:2016yiq} provides the complete boundary data on the hypersurface $t=0$ and one might be tempted
to use this boundary data and integrate just in $t$ (or $\bar{y}$).
This is possible, but does not lead to compact final expressions.
The reason is that integration in $\bar{y}$ for $\bar{x}=\mathrm{const}$ or $\xp=\mathrm{const}$ leads to polynomials of
higher degree in $\bar{y}$ in the arguments of the logarithms in eq.~(\ref{def_omega}).
We find it more convenient to first integrate in $\bar{y}$, and then in $\bar{x}$ or $\xp$, as opposed to the other way round.

We use the integration variable $\bar{x}$ for all iterated integrals not involving the square root $r_3$, while the integration
variable $\xp$ is used for iterated integrals involving the square root $r_3$.
This is unproblematic for all iterated integrals not involving trailing zeros.
For iterated integrals with trailing zeros some care has to be taken, related to the fact that
\bq
 \left. \frac{d\bar{x}}{d\xp} \right|_{\xp=0}
 & = &
 4 \left(1-\bar{y}\right).
\eq
Consider $\ln(\bar{x})$, which corresponds to
\bq
\label{example_trailing_zero}
 \int\limits_0^{\bar{x}} \frac{dx}{x}
 & = &
 \int\limits_0^{\bar{x}} d\ln\left(x\right).
\eq
Of course, strictly speaking the integral in eq.~(\ref{example_trailing_zero}) does not equal $\ln(\bar{x})$.
It is divergent due to the singularity of the integrand at the lower integration boundary.
However, it is standard practice to imply a regularisation and renormalisation procedure and to assign $\ln(\bar{x})$ to the integral
in eq.~(\ref{example_trailing_zero}).
From eq.~(\ref{rationalise_r3}) we have
\bq
 \ln\left(\bar{x}\right) 
 & = &
 2 \ln\left(2\right)
 + \ln\left(1-\bar{y}\right)
 + \ln\left(\xp\right)
 + \ln\left(1-\xp + \xp \bar{y} \right)
 \nonumber \\
 & &
 - \ln\left[ 1 + \left(1-\bar{y}+\bar{y}^2\right) \xp \right]
 - \ln\left[ 1 - \left(1-\bar{y}+\bar{y}^2\right) \xp \right].
\eq
Consider now $\omega=d\ln(\bar{x})$, which we would like to integrate on the hypersurface $\bar{y}=\mathrm{const}$ from
$\bar{x}=0$ to $\bar{x}$.
If we change variables from $\bar{x}$ to $\xp$ and integrate on the hypersurface $\bar{y}=\mathrm{const}$ from
$\xp=0$ to $\xp$ we miss the terms $2\ln(2)+\ln(1-\bar{y})$ as
\bq
 \left.
 2 d \ln\left(2\right)
 + d \ln\left(1-\bar{y}\right)
 \right|_{\bar{y}=\mathrm{const}}
 & = & 0.
\eq
We see that a change of variables as in eq.~(\ref{rationalise_r3}) implies also a change of the renormalisation prescription for iterated integrals
with trailing zeros.
Of course we would like to have a uniform prescription for all iterated integrals.
To this aim we isolate all trailing zeros in multiple polylogarithms in the variable $\xp$ in powers of logarithms $\ln(\xp)$ and make
the substitution
\bq
 \ln\left(\xp\right) 
 & \rightarrow &
 \ln\left(\xp\right) 
 + \ln\left(1-\bar{y}\right)
 + 2 \ln\left(2\right).
\eq
Alternatively, we may use instead of the variable $\xp$ a variable $x''=4(1-\bar{y})\xp$, for which
\bq
 \left. \frac{d\bar{x}}{dx''} \right|_{x''=0}
 & = &
 1.
\eq
For the integration in $\bar{y}$ we have the alphabet (with upper integration limit $\bar{y}$)
\bq
 {\mathcal A}_{\bar{y}}
 & = &
 \left\{ 0,1,2 \right\},
\eq
for the integration in $\bar{x}$ we have the alphabet (with upper integration limit $\bar{x}$)
\bq
 {\mathcal A}_{\bar{x}}
 & = &
 \left\{ 0,1,2,1+y,\frac{1+y}{y} \right\},
\eq
while for the integration in $\xp$ we have the alphabet (with upper integration limit $\xp$)
\bq
 {\mathcal A}_{\xp}
 & = &
 \left\{ 
  0,
  \frac{1}{y},
  \frac{1}{1-y+y^2},
  \frac{1}{1+y+y^2},
  - \frac{1}{1-y-y^2},
  - \frac{1}{1-y+y^2},
  \frac{1}{1+y-y^2},
  - \frac{1}{1-3y+y^2},
 \right. \nonumber \\
 & & \left.
  - \frac{1+y}{1-2y-y^3},
  \frac{1+y}{1+2y^2-y^3},
  x_1',
  x_2'
 \right\},
\eq
where $x_1'$ and $x_2'$ are the solutions for $\xp$ of the equation
\bq
  (y^4-2y^3+y^2-2y+1) \xp^2 + 2 y \xp - 1 & = & 0.
\eq
Up to weight four we obtain 44 different multiple polylogarithms from the integration in $\bar{y}$,
144 different multiple polylogarithms from the integration in $\bar{x}$
and 
4289 different multiple polylogarithms from the integration in $\xp$.
This is not surprising: The larger the alphabet, the more possibilities there are for an ordered sequence of up to four letters.

\subsection{The master integral $J_{18}$}

Up to weight four the root $r_4$ enters only the master integral $J_{18}$.
\begin{figure}
\begin{center}
\includegraphics[scale=1.0]{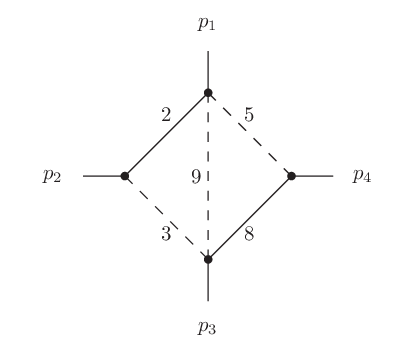}
\end{center}
\caption{
The topology for the master integral $J_{18}$.
This topology has four master integrals $J_{18}$-$J_{21}$.
Up to weight four the square root $r_4$ enters only $J_{18}$.
}
\label{fig_J18}
\end{figure}
The topology of this master integral is shown in fig.~\ref{fig_J18}.
The Feynman integral $J_{18}$ starts at ${\mathcal O}(\eps^4)$ and as we are only interested in terms up to weight four, 
we only need to compute $J_{18}^{(4)}$.
The master integral $J_{18}$ appears also as a sub-topology in the two-loop corrections for Bhabha scattering \cite{Henn:2013woa,Heller:2019gkq}.
We follow the lines of ref.~\cite{Heller:2019gkq} and compute this integral from the Feynman parametrisation
\bq
 J_{18}^{(4)}
 & = &
 4 r_4
 \int\limits_{0}^\infty d\alpha_2
 \int\limits_{0}^\infty d\alpha_5
 \int\limits_{0}^\infty d\alpha_9
 \int\limits_{0}^\infty d\alpha_3
 \int\limits_{0}^\infty d\alpha_8
 \frac{\delta\left(1-\alpha_8\right)}{{\mathcal U}{\mathcal F}},
\eq
with
\bq
 {\mathcal U}
 & = &
 \left(\alpha_2+\alpha_3\right) \left(\alpha_5+\alpha_8\right)
 + \left(\alpha_2+\alpha_3\right) \alpha_9
 + \left(\alpha_5+\alpha_8\right) \alpha_9,
 \\
 {\mathcal F}
 & = &
 \alpha_3 \alpha_5 \alpha_9 \left(-s\right)
 + \alpha_2 \alpha_8 \alpha_9 \left(2m^2-t\right)
 + \left[ \alpha_2^2 \left(\alpha_5+\alpha_8+\alpha_9\right) + \alpha_8^2 \left(\alpha_2+\alpha_3+\alpha_9\right) \right] m^2,
 \nonumber
\eq
combining the methods of linear reducibility \cite{Brown:2008}
with the algorithms for the rationalisation of square roots \cite{Besier:2018jen}.
For the former we use the program ``HyperInt'' \cite{Panzer:2014caa}, for the latter the program `RationalizeRoots'' \cite{Besier:2019kco}.
This allows us to express $J_{18}^{(4)}$ in terms of multiple polylogarithms.
We obtain an alphabet with $23$ letters and an expression for $J_{18}^{(4)}$ in terms of $2330$ different multiple polylogarithms.

% -----------------------------------------------------------------------------

\section{Results}
\label{sect:results}

Albeit the fact that the result for the scalar double box integral 
$J_{23}$ is rather compact,
\bq
 J_{23}
 & = &
 4 \left[ G\left(1,1,1;\bar{y}\right) + \zeta_2 G\left(1;\bar{y}\right) \right] \eps^3
 + 4 \left\{
           2 G\left(2,1,1,1;\bar{y}\right)
           + 2 G\left(0,1,1,1;\bar{y}\right)
           - 2 G\left(1,1,2,1;\bar{y}\right)
 \right. \nonumber \\
 & & \left.
           - 2 G\left(1,1,0,1;\bar{y}\right)
           + 2 \zeta_2 \left[ G\left(2,1;\bar{y}\right) + G\left(0,1;\bar{y}\right) - G\left(1,1;\bar{y}\right) \right]
           - \zeta_3 G\left(1;\bar{y}\right) 
 \right. \nonumber \\
 & & \left.
           + 2 \left[ G\left(1,1,1;\bar{y}\right) + \zeta_2 G\left(1;\bar{y}\right) \right] \left[ G\left(1,\bar{x}\right) -2 \ln\left(\bar{x}\right) \right] 
   \right\} \eps^4
 + {\mathcal O}\left(\eps^5\right),
\eq
some of the other master integrals have rather involved expressions in terms of multiple polylogarithms.
For this reason we give the results in electronic form.
On the one hand, we provide symbolic expressions in terms of multiple polylogarithms for all master integrals
up to weight four.
On the other hand, we provide a numerical program, which evaluates all master integrals up to weight four
at a given kinematic point inside the physical region with the help of the numerical evaluation routines for multiple polylogarithms of
\verb|GiNaC| \cite{Bauer:2000cp,Vollinga:2004sn}.
The files are described in more detail in appendix~\ref{sect:supplement}.

Physical regions in the kinematic space are (we always assume $m^2>0$)
\begin{description}
\item{Region I}: $s<0$, $t<0$, $u>4m^2$,
\item{Region II}: $s<0$, $t>4m^2$, $u<0$,
\item{Region III}: $s>4m^2$, $t<0$, $u<0$.
\end{description}
Region I is the Euclidean region, region II is the one relevant to the inspiral process of a binary system,
region III corresponds in particle physics to the annihilation-creation process.
We first derive the result in the Euclidean region.
The result may be analytically continued to other regions.
The analytic continuation can be done by giving the variables $s$ and $t$ 
a small imaginary part according to Feynman's $i\delta$-prescription
\bq
\label{Feynman_prescription}
 s \; \rightarrow \; s + i \delta_s,
 & &
 t \; \rightarrow \; t + i \delta_t,
 \;\;\;\;\;\;
 \delta_s, \delta_t > 0,
\eq
provided the following two conditions hold:
There is a continuous path in kinematic space from the Euclidean region to the kinematic point of interest, such that
\begin{enumerate}
\item no branch cut of the square roots is crossed,
\item no branch cut of the multiple polylogarithms is crossed.
\end{enumerate}
Requirement $1$ is rather easy to satisfy for all real values of $s$ and $t$:
We use the form of the square roots as in eq.~(\ref{def_roots}).
The replacement in eq.~(\ref{Feynman_prescription}) selects the correct branch of all square roots except possibly
the square root
\bq
 \sqrt{4m^6-s\left(m^2-t\right)^2}.
\eq
For $s>4m^6/(m^2-t)^2$ the argument of the square root is negative
and the correct branch of the square root is selected by the imaginary part of $s$.
The kinematic variable $t$ enters only in the combination $(m^2-t)^2$ and a possible small imaginary part of $t$ is not relevant for the
selection of the branch cut.
Thus we set in the case $s>4m^6/(m^2-t)^2$ and $0<t<m^2$
\bq
 \delta_t & < & \frac{\left(m^2-t\right)}{2s} \delta_s.
\eq
This ensures that the small imaginary part of $s$ dominates over the small imaginary part of $t$.
In all other regions we set $\delta_s = \delta_t$.

Requirement 2 is more subtle. For all multiple polylogarithms we extract trailing zeros and then normalise
the upper integration limit in the multiple polylogarithms to one.
Thus requirement 2 translates to the requirements that no argument of an explicit logarithm 
(obtained from extracting trailing zeros) crosses the negative
real axis and no letter of a multiple polylogarithm crosses the line segment $[0,1]$.

If a crossing occurs and the final value is within an infinitesimal distance from the branch cut, we may try to rescue
the situation by modifying the relative size of $\delta_s$ and $\delta_t$.
If this is not possible or if the final value is a finite distance away from the branch cut, we have to compensate
the branch cut crossing by adding the corresponding monodromy.

We have scanned several kinematic points and found that a branch cut crossing of explicit logarithms or 
multiple polylogarithms occurs only in the unphysical region $s>0$ and $t>0$.
As our main interest are the physical regions I, II and III, our program implements the analytic continuation
as in eq.~(\ref{Feynman_prescription}) with $\delta_s=\delta_t$.

As a reference point we give here numerical results for the point
\bq
\label{kinematic_point}
 s \; = \; - \frac{1}{36} \; \mathrm{GeV}^2,
 \;\;\;\;\;\;
 t \; = \; 5 \; \mathrm{GeV}^2,
 \;\;\;\;\;\;
 m^2 \; = \; 1 \; \mathrm{GeV}^2.
\eq
This is a point from region II.
We set $\mu^2=m^2$.
\begin{table}[!htbp]
\begin{center}
{\scriptsize
\begin{tabular}{|l|lllll|}
 \hline 
 & $\eps^0$ & $\eps^1$ & $\eps^2$ & $\eps^3$ & $\eps^4$ \\
 \hline 
$J_{ 1}$ & $        1$ & $        0$ & $ 1.6449341$ & $-0.80137127$ & $ 1.8940657$ \\ 
$J_{ 2}$ & $       -1$ & $-3.5835189$ & $-6.420804$ & $-4.4642058$ & $ 7.8627655$ \\ 
$J_{ 3}$ & $     -0.5$ & $        0$ & $-4.1123352$ & $-4.407542$ & $-40.992992$ \\ 
$J_{ 4}$ & $        1$ & $ 7.1670379$ & $ 24.038282$ & $ 36.746282$ & $-39.627704$ \\ 
$J_{ 5}$ & $        0$ & $-0.96242365 +  3.1415927 i$ & $-23.544931 -5.0561983 i$ & $-18.55474 -71.841236 i$ & $ 92.319226 -171.0195 i$ \\ 
$J_{ 6}$ & $        0$ & $        0$ & $-8.9433451 -6.0470861 i$ & $ 10.268808 -29.261043 i$ & $ 36.85189 -50.345329 i$ \\ 
$J_{ 7}$ & $        1$ & $ 7.1670379$ & $ 24.038282$ & $ 43.958624$ & $ 21.804329$ \\ 
$J_{ 8}$ & $        0$ & $        0$ & $ 8.9398354$ & $ 48.985113$ & $ 168.17193$ \\ 
$J_{ 9}$ & $        0$ & $        0$ & $-17.879671$ & $-168.99678$ & $-962.19027$ \\ 
$J_{ 10}$ & $        0$ & $        0$ & $        0$ & $ 3.4772053$ & $ 19.173012$ \\ 
$J_{ 11}$ & $        0$ & $        0$ & $ 4.4699177$ & $ 45.726399$ & $ 286.64965$ \\ 
$J_{ 12}$ & $     0.25$ & $        0$ & $-2.7991685$ & $-3.7744625$ & $ 40.39835$ \\ 
$J_{ 13}$ & $        0$ & $        0$ & $        0$ & $-24.14137 -11.639717 i$ & $-86.425957 -104.64508 i$ \\ 
$J_{ 14}$ & $        0$ & $        0$ & $-8.9398354$ & $-81.021183$ & $-386.4065$ \\ 
$J_{ 15}$ & $        0$ & $        0$ & $        0$ & $ 31.53239 +  23.957578 i$ & $ 274.64809 +  294.78346 i$ \\ 
$J_{ 16}$ & $        0$ & $        0$ & $        0$ & $-31.553588 -24.931446 i$ & $-270.62397 -299.81333 i$ \\ 
$J_{ 17}$ & $        0$ & $-0.48121183 +  1.5707963 i$ & $-7.3256456 +  8.7298576 i$ & $-26.79551 +  15.696906 i$ & $ 17.270388 -14.29595 i$ \\ 
$J_{ 18}$ & $        0$ & $        0$ & $        0$ & $        0$ & $-221.70403 -52.195623 i$ \\ 
$J_{ 19}$ & $        0$ & $ 1.9248473 -6.2831853 i$ & $ 29.302582 -34.91943 i$ & $ 131.32341 -51.147906 i$ & $ 62.917726 +  164.00682 i$ \\ 
$J_{ 20}$ & $        0$ & $        0$ & $-8.9398354$ & $-56.904487 +  49.862893 i$ & $-167.11741 +  635.25378 i$ \\ 
$J_{ 21}$ & $      0.5$ & $ 3.5835189$ & $ 9.1924025 -3.0235431 i$ & $ 8.1915237 -23.451378 i$ & $ 60.572872 -56.78219 i$ \\ 
$J_{ 22}$ & $        0$ & $        0$ & $        0$ & $        0$ & $ 79.920657$ \\ 
$J_{ 23}$ & $        0$ & $        0$ & $        0$ & $ 12.070685 +  5.8198587 i$ & $ 101.45906 +  57.936199 i$ \\ 
$J_{ 24}$ & $        0$ & $        0$ & $        0$ & $        0$ & $-47.316121 -24.238497 i$ \\ 
$J_{ 25}$ & $      0.5$ & $ 3.5835189$ & $ 12.019141$ & $ 18.373141$ & $ 88.684345 +  7.6301454 i$ \\ 
 \hline 
\end{tabular}
}
\end{center}
\caption{
Numerical results for the first five terms of the $\eps$-expansion of the master integrals $J_{1}$-$J_{25}$ 
at the kinematic point of eq.~(\ref{kinematic_point}).
}
\label{table_numerical_results}
\end{table}
The values of the master integrals at this point are given to $8$ digits in table~\ref{table_numerical_results}.
In addition we verified the first few digits at several other kinematic points
with the help of the programs \verb|sector_decomposition| \cite{Bogner:2007cr}
and
\verb|pySecDec| \cite{Borowka:2017idc,Hahn:2004fe,Ruijl:2017dtg,GSL}.

% -----------------------------------------------------------------------------

\section{Conclusions}
\label{sect:conclusions}

In this article we presented the results for all master integrals associated to the two-loop H-graph with equal masses
and up to weight four
in terms of multiple polylogarithms.
The challenge in obtaining this result is the occurrence of four square roots in the differential equations
for the master integrals.
Although we cannot rationalise simultaneously all square roots, we were nevertheless able to express all master integrals
up to weight four in terms of multiple polylogarithms.
The techniques we used carry over to more complicated Feynman integrals, in particular the H-graph with unequal masses
can be treated along the same lines.

\subsubsection*{Note added}

A detailed comparison with \verb|AMFlow| \cite{Liu:2022chg} showed that the boundary value for $J^{(4)}_{25}$ should
be $-\frac{57}{8} \zeta_4$ instead of $-7\zeta_4$, correcting eq.~(4.4) of ref.~\cite{Bianchi:2016yiq}.
For this particular integral the small difference is hard to detect with numerical programs 
based on sector decomposition.
In the version \verb|arXiv:2104.07488v3| we corrected the numerical value of $J^{(4)}_{25}$ in table~\ref{table_numerical_results} and the data in the supplementary material.
We updated the ancillary file to \verb|hequal-1.0.1.tar.gz|.

% -----------------------------------------------------------------------------

\begin{appendix}

\section{Supplementary material}
\label{sect:supplement}

Attached to the arxiv version of this article is an electronic file 
\verb|hequal-1.0.1.tar.gz|.
This file contains symbolic expressions in terms of multiple polylogarithms for all master integrals
and a numerical program to evaluate all master integrals up to weight four
at a given kinematic point in the physical region.

After unpacking, the symbolic expressions can be found in the file 
\begin{center}
 \verb|supplementary_material.mpl|
\end{center}
in the \verb|maple_files|-directory.
The file \verb|supplementary_material.mpl|
is in ASCII format with {\tt Maple} syntax, defining the quantities
\begin{center}
 \verb|A|, \; \verb|log_lst|, \; \verb|letter_lst_ybar|, \; \verb|letter_lst_xbar|, \; \verb|letter_lst_xp|, \; \verb|letter_lst_J18|, \; \verb|J|.
\end{center}
The matrix \verb|A| appears in the differential equation eq.~(\ref{def_diff_eq})
\bq
 d \vec{J} & = & A \vec{J}.
\eq
The entries of the matrix $A$ are linear combinations of $\omega_1$, ..., $\omega_{17}$, defined in eq.~(\ref{def_omega}).
These differential forms are denoted by
\begin{center}
 \verb|omega_1|, ..., \verb|omega_17|.
\end{center}
The dimensional regularisation parameter $\eps$ is denoted by \verb|eps|.
The variables $s$, $t$, $y$, $\bar{y}$, $\bar{x}$, $\xp$, $x_1'$ and $x_2'$ are denoted by
\begin{center}
 \verb|s|, \verb|t|, \verb|y|, \verb|ybar|, \verb|xbar|, \verb|xp|, \verb|xp_r1|, \verb|xp_r2|,
\end{center}
respectively.
The square roots are denoted by \verb|r1|, \verb|r2|, \verb|r3| and \verb|r4|.
The expression for $J_{18}$ involves two additional roots, which are denoted by \verb|r7| and \verb|r8|
and defined by
\bq
 r_7 \; = \; \sqrt{-s}\sqrt{-t},
 & &
 r_8 \; = \; \sqrt{4 m^4-st}.
\eq
The zeta values $\zeta_2$, $\zeta_3$, $\zeta_4$ are denoted by
\begin{center}
 \verb|zeta_2|, \verb|zeta_3|, \verb|zeta_4|.
\end{center}
The lists
\begin{center}
 \verb|log_lst|, \; \verb|letter_lst_ybar|, \; \verb|letter_lst_xbar|, \; \verb|letter_lst_xp|, \; \verb|letter_lst_J18|
\end{center}
contain the definitions for single logarithms and the letters of the various alphabets.
The vector \verb|J| contains the results for the master integrals up to order $\eps^4$ in terms of multiple polylogarithms.
For the notation of multiple polylogarithms we give an example:
$G(l_1,l_2,l_3;1)$ is denoted by
\begin{center}
 \verb|Glog([l_1,l_2,l_3],1)|.
\end{center}

The numerical program
to evaluate all master integrals up to weight four
at a given kinematic point requires the \verb|GiNaC| library to be installed.
Running the commands
\begin{verbatim}
 ./configure
 make
 cd bin
 ./hequal
\end{verbatim}
will compile and run the numerical program \verb|hequal| in the \verb|bin| directory.
The user may modify the variables
\verb|Digits|, \verb|s|, \verb|t|, \verb|m2| in \verb|hequal.cc|.
The variable \verb|m2| denotes the mass squared $m^2$.

\end{appendix}

% ----------------------------------------------
% references
{\footnotesize
\bibliography{/home/stefanw/notes/biblio}
\bibliographystyle{/home/stefanw/latex-style/h-physrev5}
}

\end{document}